\documentstyle [12pt]{article}
\begin{document}
\baselineskip .3in
\begin{titlepage}
\begin{center}{\large {\bf Statistical mechanics of money: How saving propensity affects its distribution}}
\vskip .2in
{\bf Anirban Chakraborti}$~^{(1)}$ and
{\bf Bikas K. Chakrabarti}$~^{(2)}$\\
{\it Saha Institute of Nuclear Physics},\\
{\it 1/AF Bidhan Nagar, Calcutta 700 064, India.}\\
\end{center}
\vskip .3in
{\bf Abstract}\\
\noindent
We consider a simple model of a closed economic system where the total money is 
conserved and the number of economic agents is fixed. Analogous to statistical
systems in equilibrium, money and the average money per economic agent are 
equivalent to energy and temperature, respectively. We investigate the effect 
of the saving propensity of the agents on the stationary or equilibrium 
probability distribution of money. 
When the agents do not save, the equilibrium  money distribution becomes the 
usual Gibb's distribution, characteristic of non-interacting agents.  
However with saving, even for individual self-interest, the 
dynamics becomes cooperative and the resulting asymmetric Gaussian-like 
stationary distribution acquires global ordering properties. 
Intriguing singularities are observed in the stationary money 
distribution in the market, as functions of the marginal saving propensity 
of the agents.
\end{titlepage}
\newpage
\noindent
{\bf 1 Introduction}\\
\noindent
The interacting dynamical nature of any economic sector composed of many 
cooperatively interacting agents, has many features in common with the 
statistical physics of interacting systems. In fact, economists like Pareto 
investigated the power law properties of the wealth distributions more than a
century ago and Stigler studied the market fluctuations by employing 
Monte Carlo methods more than thirty years back [1].
Motivated by these investigations, many
 efforts are being made recently to apply the statistical physics methods to
various economic problems. A
major part of the recent efforts has gone in investigating the nature of 
fluctuations and their distributions in the stock markets [2]. 
We believe, however, a fundamental dynamics occurs in the money market, which
 affects 
strongly the dynamics of other sectors in the economy. An understanding of the 
statistical mechanics of the money market is therefore essential, and some of 
its features are very intriguing [3, 4]. Dragulescu and 
Yakovenko [5] have shown very recently that for any arbitrary and random sharing
 but locally 
conserving money transaction between any two agents in a market, the money 
distribution goes to the equilibrium Gibb's distribution of  statistical 
mechanics: $P(m)=(1/T)\exp (-m/T)$ where $T=M/N$, the average money per 
agent in the market ($M$ is the total money of $N$ agents in the market). 
This equilibrium distribution is extremely robust and various kinds of 
inter-agent monetary transactions, which locally conserve the total money, lead
to the Gibb's probability distribution satisfying $P(m_1)P(m_2)=P(m_1+m_2)$. A 
major achievement of this study [5] has been the precise identification of the 
temperature $T$ as the average money per agent in the market. This is due to 
the fact that the probability distribution is normalized and the total money is 
conserved. This precise identification had been missing in 
many of the previous attempts [3, 4] though the identification from fiscal 
policy considerations was indeed very close [3].
They had considered a market consisting of $N$ agents, where 
initially each one gets any arbitrary share $m_i$ of the total money $M$ in the
market ($\sum_{i=1}^N m_i=M$; $M$ and $N$ fixed). 
The ``trade'' dynamics goes as follows. Select any 
two arbitrary agents $i$ and $j$ with money $m_i$ and $m_j$, respectively. 
These two agents then exchange their money through some trade, keeping their
total amount of money $m_i+m_j$ conserved and no debt is allowed ($m_i\geq 0$ at
any stage of the trade). There is no other restriction in the trade. 
Extensive numerical simulations show that this and various modifications of 
trade, like multi-agent transactions, etc., all lead to the robust Gibb's 
distribution, independent of the initial distribution the market starts with. 
So, most of the agents end-up in this market with very little money! Supply 
of more money in the market (increasing $T$) can increase the width of the 
distribution, but the most probable money for any agent in the market remains 
zero. It may be mentioned that Ispolatov {\it et al.} had studied earlier the 
(non-equilibrium) wealth distributions in various asset exchange models [6] 
where the trade dynamics do not have time-reversal symmetry. Of course, the 
study of Levy and Solomon [7] indicates a power law distribution for the wealth
in realistic (possibly non-equilibrium) markets or societies.

Although the model of Dragulescu and Yakovenko [5] is a simple and interesting 
one to start with, a very natural ingredient
for any realistic economic agent is missing in the model: no economic agent
trades with the entire money he or she possesses without saving a part of it;
saving propensity is too natural a tendency for any economic agent [8].
This saving propensity of course varies from agent to agent and even with the 
accumulation of wealth of a single agent. There are also country-wise variations
of this saving propensity. We note that the fraction of savings $\lambda$, 
called the ``marginal propensity to save'' by economists [8], remains 
fairly constant, independent of the agents. We have taken it to be a constant 
in the model considered here. We show that in presence of 
this saving factor $\lambda (\neq 0)$ the money market becomes truly cooperative
in nature and ``critical behaviour'' [9] sets in. The multiplicative property 
of the Gibb's distribution $P(m)$, discussed above,
was responsible for the absence of any cooperative feature of the 
statistics for $\lambda =0$. Once the local or individual measure of saving 
propensity is introduced  ($\lambda \neq 0$), a global order 
emerges in the entire money market, giving a non-vanishing most probable
money for each agent in the market and nontrivial critical behaviour of the 
resulting statistics in the equilibrium.\\
\noindent
{\bf 2 Model and simulation results for money distribution}\\
\noindent
We again consider a simple model of the closed economic system where the total 
amount of money $M$ is conserved and the number of economic agents $N$ is 
fixed. 
Each economic agent $i$, which may be an individual or a corporate, possesses
money $m_i$. An economic agent can exchange money with any other agent through 
some trade, keeping the total amount of money of both the agents conserved.
We assume that each economic agent saves a fraction $\lambda$ of its money 
$m_i$ before trading. We again assume that an agent's money must always be 
non-negative and therefore no debt is permitted.
Let us now consider that an arbitrary pair of agents $i$ and $j$ get engaged in
a trade so that their money $m_i$ and $m_j$ change by amounts $\Delta m_i$ and  
$\Delta m_j$ to become $m_i^{\prime }$ and $m_j^{\prime }$, where $\Delta m_i$ 
is a random fraction of $(1-\lambda)(m_i + m_j)$ and $\Delta m_j$ is the rest of
it. Conservation of the total money in each trade is ensured, as earlier.

We performed computer simulations with fixed number of agents $N$ and total 
money $M=NT$. Most of our simulation results are for $N=500$ and $T=100$.
However, we checked results for different $N$ values ($250\leq N\leq 1000$) to
check finite size effects, etc., and with different $T$ values (upto $10000$).
Initially we divided the total money $M$ amongst $N$ agents equally so that 
$m_i=M/N=T$ for all $i$. We chose a fixed value of $\lambda $ between zero and 
unity and investigated its effects on the equilibrium distribution of money
$P(m)$ in the market, giving the (normalized) number of agents $P$ with money
$m$. We choose randomly two agents $i$ and $j$ having money $m_i$ and $m_j$, 
respectively. Then $\Delta m_i=\epsilon (1-\lambda)(m_i + m_j)$
and  $\Delta m_j=(1-\epsilon ) (1-\lambda)(m_i + m_j)$, where $\epsilon$ is a 
random number between zero and unity.  
Then $m_i^{\prime }=\lambda m_i+\Delta m_i$ and 
$m_j^{\prime }=\lambda m_j+\Delta m_j$ after the trade. 
Alternatively, this trade can also be viewed as 
$m_i\rightarrow {m_i}^{\prime }$,
$m_j \rightarrow {m_j}^{\prime }$ where ${m_i}^{\prime }=m_i-\Delta m$,
${m_j}^{\prime }=m_j+\Delta m$ with 
$\Delta m=(1-\lambda)[m_i- \epsilon (m_i+m_j)]$.
This can be checked by straightforward substitution.
These 
trades were repeated for large number of arbitrary choices of pairs of agents;
each of these random choices of pairs is considered as one trade and 
consequently one time unit. The typical time upto which we run this algorithm 
is above $5000$ (sometimes upto $50000$ for larger
$\lambda$). The probability distribution was determined after every $50$
time steps till a stationary distribution was obtained. We then took an average
over $2000$ such stationary distributions to obtain a smooth distribution.
We checked that the equilibrium distribution obtained is again extremely robust
and does not depend at all on the initial distribution of money in the market.
We investigated the nature of the equilibrium distribution $P(m)$ for various
values of $\lambda$ ($0\leq \lambda <1$). Apart from the stationary 
distribution, we also 
investigated the ``relaxation'' behaviour [9] of the distributions and obtained 
the time variations of $P_1\equiv P(m=T)$ till a steady behaviour was found.
We define the relaxation time $\tau_R$ as the earliest time where $P_1(t)$
becomes practically independent of time.

The results for the equilibrium distribution $P(m)$ are shown
in Fig. 1, for some values of $\lambda$. The inset shows that the
equilibrium distribution is independent of the market size $N$ and the 
average money in the market or temperature $T$. 
The real money exchanged randomly in any trade is less than the total money,
because of the saving by each agent. This destroys the multiplicative property
of the distribution $P(m)$ (seen earlier for $\lambda =0$) and $P(m)$ changes 
from the Gibb's form to the asymmetric Gaussian-like form as soon as a finite
$\lambda$ is introduced. The $\lambda =0$ case was practically a random noise 
dominating one and therefore effectively a non-interacting market. Introduction
of a finite amount of saving ($\lambda \neq 0$), dictated by individual
self-interest, immediately makes the money dynamics cooperative and the global 
ordering (in the distribution) is achieved. This kind of self-organization in 
the money market, coming out of pure self-interest of each agent, is reminiscent
of the ``invisible hand'' effect [8, 10] in the ``free market'' suggested 
originally by Adam Smith in 1776.

The relaxation behaviour of the distribution is shown in Fig. 2, where the time
variation of $P_1(t)$ for different values of $\lambda$ is shown. Since 
$m_i/T=1$ for all $i$ to start with, 
 $P_1(t)$ starts falling from unity in all the cases (not shown in Fig. 2). 
After some initial rapid decay, we see an extremely slow spin-glass type 
[4, 9] ($\ln ~ t$) relaxation behaviour.
The inset shows the typical variation of the relaxation time $\tau_R$ with 
$\lambda$. The dynamics obviously becomes slower with increasing $\lambda$ and 
$\tau_R$ seems to diverge as $\lambda$ approaches unity. Precisely at 
$\lambda=1$, the dynamics of course stops.

An important feature of this humped distribution $P(m)$ at
any non-vanishing $\lambda$ is the variation of the most probable money 
$m_p(\lambda )$ ( where $P(m)$ becomes maximum) of the agents. 
As discussed before, $m_p=0$ for $\lambda =0$ (Gibb's distribution) and most of
 the economic agents
in the market end-up losing most of their money. However, even with pure 
self-interest of each agent for saving a factor $\lambda$ of {\it its} own money
 at any trade, a global feature emerges: the
entire market ends-up with a most-probable money $m_p(\lambda )$.
This $m_p(\lambda )$ shifts in an interesting manner from $m_p=0$ (for 
$\lambda =0$) to $m_p \rightarrow T$ (for 
$\lambda \rightarrow 1$). We find that initially, for small  $\lambda$, 
$m_p(\lambda )$ varies very 
closely as $\lambda ^{1/2}$ and then it crosses over at  
$\lambda ={\lambda}_c\simeq 0.45$ to  $\lambda ^{1/3}$ variation. This is shown 
in Fig. 3, where the 
curves for  $\lambda ^{1/2}$ and  $\lambda ^{1/3}$ are also indicated.
We do not have any idea why this crossover in the exponent for  $\lambda $
occurs at any finite  $\lambda ~(\simeq 0.45)$. Also the exponents $1/2$ and 
$1/3$ are very intriguing. We checked that this crossover point is perhaps also
the point where the probability of an agent having zero money 
 $P_0\equiv P(m=0)$ just disappears, as shown in the inset of Fig. 3.
The half-width $\Delta m_{p}$ and the peak height 
$P_{ m_{p}}$ of the equilibrium distribution, which scale practically as 
$(1-\lambda )^{1/2}$ and $(1-\lambda )^{-1/2}$ respectively (see Fig. 4), do not
have any irregularity there.    
In fact, no other property of the equilibrium distribution $P(m)$ has any 
irregularity at ${\lambda}_c$. It may be mentioned that Ispolatov {\it et al.}
also observed a singularity in the power law of non-equilibrium growth 
distribution of wealth in their multiplicative asset exchange model having
broken time-reversal symmetric trade dynamics.

We noted that although the total distribution assumes some global cooperative 
feature, each individual's money $m_i$ fluctuates randomly.
Fig. 5 shows the time variation of the money of an arbitrarily chosen individual
 in the market for two different values of  $\lambda $. The inset shows the 
variation of the time-averaged money $\langle m_i \rangle$ of the agent and its
fluctuation 
$\Delta m_i\equiv \sqrt{\langle (m_i-\langle m_i \rangle )^2 \rangle }$ with
 $\lambda$ after relaxation ($t>\tau_R $). Since the total money is conserved,
 $\langle m_i \rangle$ remains constant ($=T$) here, while $\Delta m_i$ goes 
down 
with  $\lambda$ as $(1- \lambda )$. This is because at any time the agents keep
a fixed fraction of their individual money and receive a random fraction 
of the money traded proportional to $(1- \lambda )$.

\noindent
{\bf 3 Summary and discussion}\\
\noindent
We thus considered here a very simple model of money market where the total 
money $M$ and 
the number of agents $N$ is fixed. Each pair of arbitrarily chosen agents in the
 market exchange money with each other through a trade (each trade considered as one 
time unit). During the trade, each agent saves a fraction  $\lambda$ of its own
money at that time and exchanges randomly out of the remaining money,
conserving the total amount of money and not allowing any debt. We find that 
for  $\lambda =0$, the market effectively becomes non-interacting and, no matter 
what initial distribution the agents start with, the resulting money 
distribution becomes the equilibrium Gibb's distribution [5], where most of the 
agents in the market end-up with very little money. For any non-vanishing  
$\lambda$,
the equilibrium distribution becomes the asymmetric Gaussian-like with the 
most probable money $m_p$ (corresponding to the peak in $P(m)$) shifting away
from $m=0$ with increasing  $\lambda$. This global feature, coming out
of the individual self-interest of saving a part of its own money, can be
considered to be a demonstration of the self-organization in the market 
suggested long ago by Adam Smith (``invisible hand'' [8, 10]). Apart 
from this we find 
intriguing singularities appearing in the equilibrium distribution $P(m)$: 
$m_p\sim  {\lambda }^{\nu}$ where $\nu =1/2$ for  
$\lambda <  {\lambda }_c~(\simeq 0.45)$ and  $\nu =1/3$ for 
${\lambda }_c<\lambda <1$. Also, $\Delta m_p\sim (1-\lambda)^{1/2}$ and
$P_{m_p}\sim (1-\lambda)^{-1/2}$. 

These singularities in the equilibrium distribution come obviously from the
cooperative nature of the market interactions induced by the saving propensity 
of the agents. It may be mentioned that while such singular behaviour in the 
equilibrium money distribution is very natural here, in the corresponding 
physical (gas) system of Newtonian particles one gets regular distributions,
e.g., the Gibb's distribution 
(or for that matter, Bose or Fermi distributions for quantum particles)
 and never any singularity, because of the absence of any physical equivalent
of the ``saving'' factor
there. We thus believe, while regular distributions 
are common for minimally interacting physical many-body systems in
equilibrium, singular distributions are typical of any 
working model of the markets.\\
\vskip .1in
\noindent
{\bf Acknowledgements: } We are grateful to A. Dutta, S. S. Manna, S. Pradhan 
and P. Sen for useful discussions, and S. Redner for some useful 
communications. We also thank the anonymous referee for suggesting important
corrections and improvement of the manuscript.\\
\newpage
\noindent
{\bf References}\\
\vskip .1in
\noindent
{\it e-mail addresses} :

\noindent
$^{(1)}$anirban@cmp.saha.ernet.in

\noindent
$^{(2)}$bikas@cmp.saha.ernet.in
\vskip .2 in

\noindent
1. V. Pareto, {\it Cours d'Economie Politique}, Lausanne and Paris (1897);
G. J. Stigler, {\it J. Business}, {\bf 37}, 117 (1964).\\
\noindent
2. See e.g., R. N. Mantegna and H. E. Stanley, {\it An Introduction to 
Econophysics}, Cambridge University Press, Cambridge (2000).\\
\noindent
3. B. K. Chakrabarti and S. Marjit, {\it Ind. J. Phys.}, {\bf 69 B}, 681 (1995).\\
\noindent
4. S. Moss de Oliveira, P. M. C. de Oliveira and D. Stauffer, {\it Evolution, 
Money, War and
Computers}, B. G. Teubner, Stuttgart (1999); P. W. Anderson, K. J. Arrow and D.
 Pines (Eds), {\it The Economy as an Evolving Complex system}, Addison-Wesley, 
Redwood City (1988).\\
\noindent
5. A. Dragulescu and V. M. Yakovenko, {\it cond-mat/0001432}.\\ 
\noindent
6. S. Ispolatov, P. L. Krapivsky and S. Redner, {\it Euro. Phys. J. B}, 
{\bf 2}, 267-276 (1998).\\
\noindent
7. M. Levy and S. Solomon, {\it Physica A}, {\bf 242}, 90 (1997).\\ 
\noindent
8. P. A. Samuelson, {\it Economics}, $11$th Edition, McGraw-Hill Inc., Auckland,
 195-208 (1980).\\ 
\noindent
9. M. Plischke and B. Bergersen, {\it Equilibrium Statistical Physics}, 2nd 
Edition, World Scientific, Singapore (1994); S. K. Ma, {\it Statistical
Mechanics}, World Scientific, Singapore (1985).\\ 
\noindent
10. See e.g., J. M. Keynes, {\it The General Theory of Employment, Interest and
Money}, The Royal Economic Society, Macmillan Press, London (1973).\\ 

\newpage
\noindent
{\bf Figure captions}\\
\vskip .1 in
\noindent
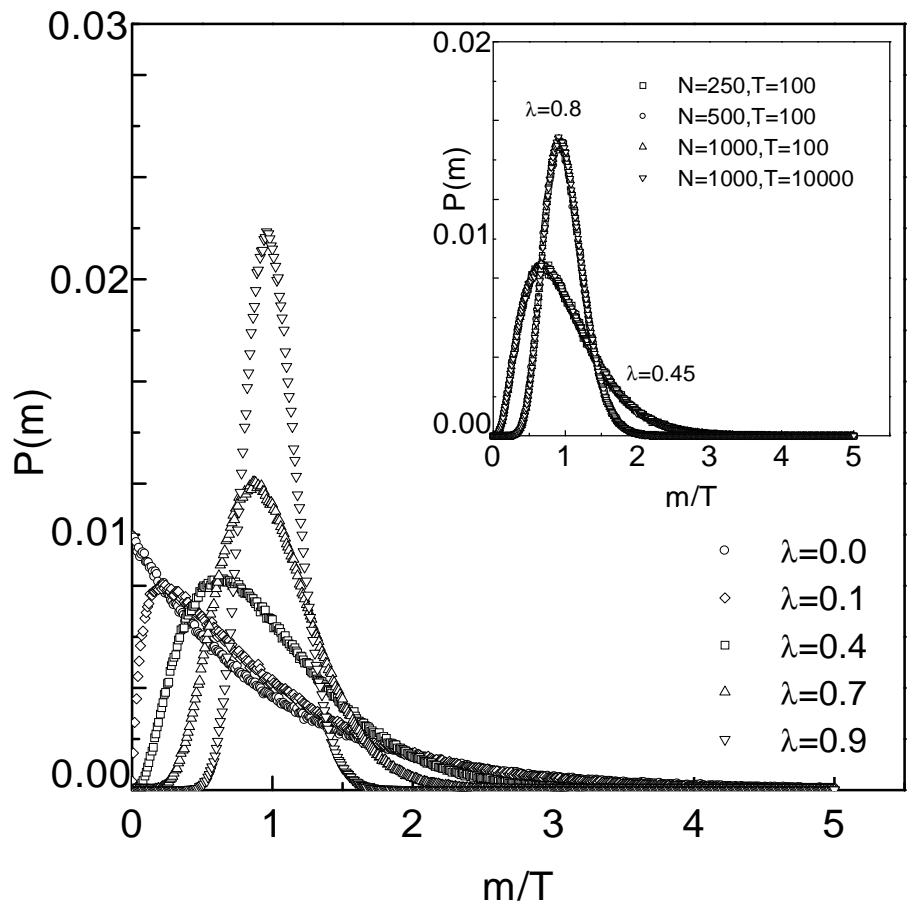
{\bf Fig. 1} : The stationary money distribution $P(m)$ versus money $m$ for 
different saving propensity factor $\lambda$ ($N=500,T=100$). The inset shows 
the same for different $N$ and $T$ at two fixed values of $\lambda$. The 
superposition of data indicates the absence of finite size effects and the 
temperature independence.\\
\noindent
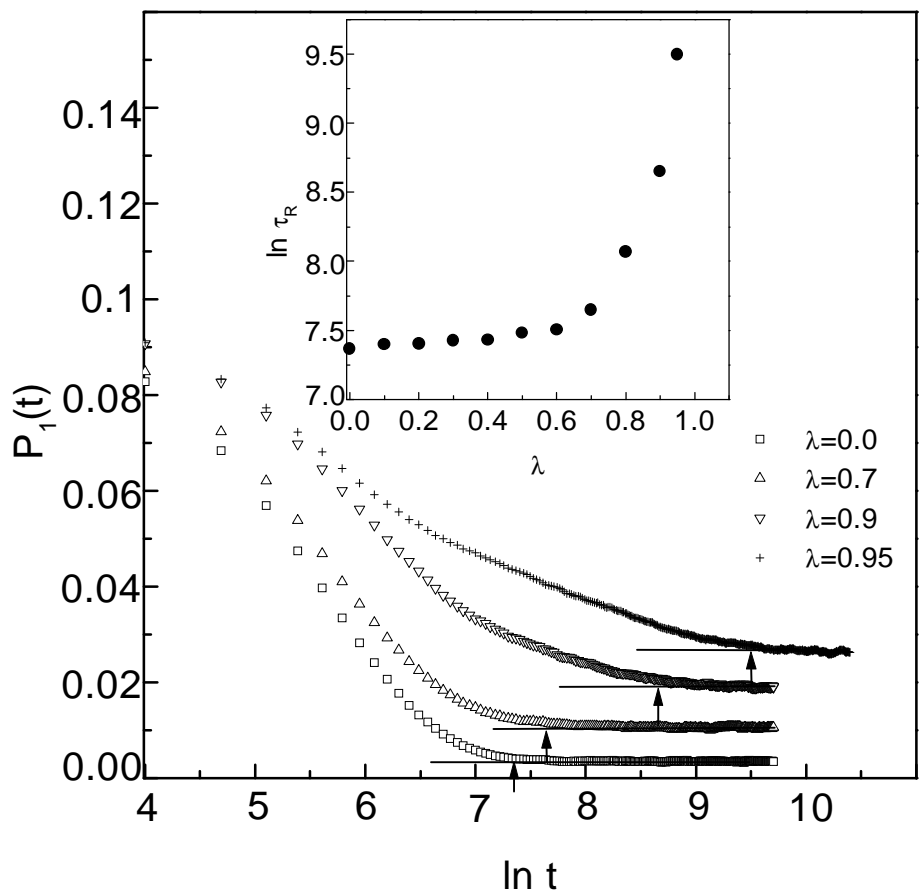
{\bf Fig. 2} : Relaxation of the distribution: $P_1(t)$ versus the time $t$ 
(number of trades) for different $\lambda$. Because of extremely slow 
relaxation, we show in the $\ln ~t$ scale for the later stage only. The vertical
 arrows 
 indicate the relaxation time $\tau_R$. The inset shows the typical variation of
 $\tau_R$ with $\lambda$; $\tau_R$ diverges as $\lambda \rightarrow 1$.\\ 
\noindent
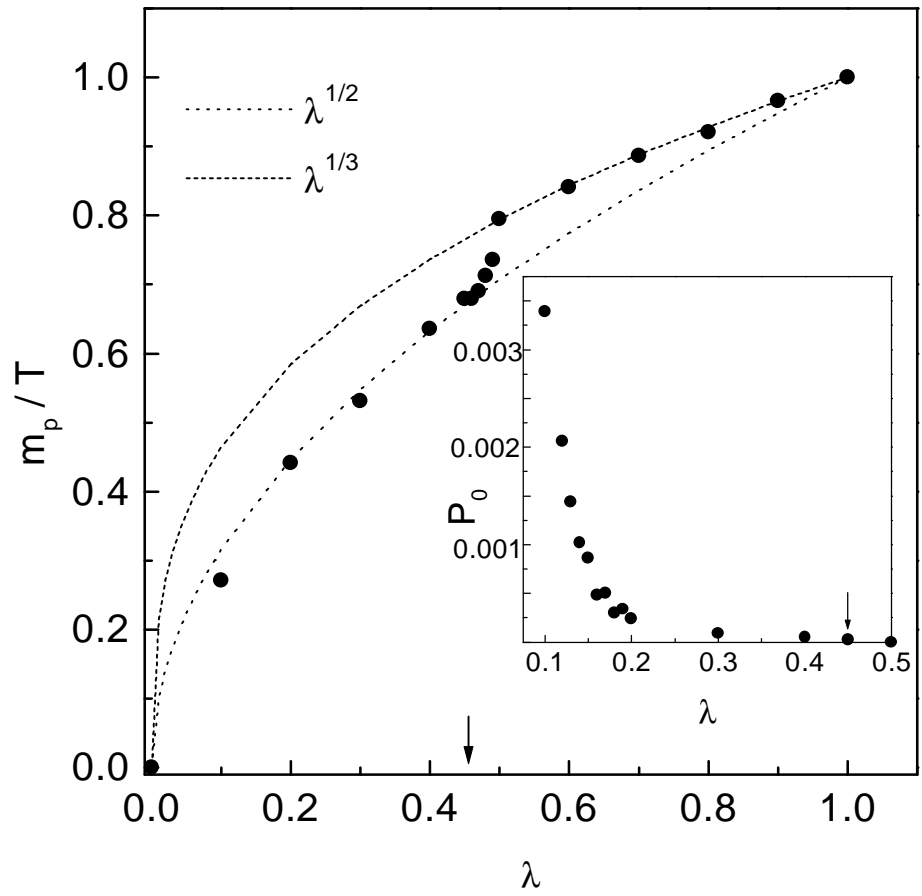
{\bf Fig. 3} : The variation of the most probable money $m_p$ of an agent (peak
 position of the equilibrium distribution $P(m)$) as a function of $\lambda$. 
The dotted and dashed curves correspond to ${\lambda }^{1/2}$ and 
${\lambda }^{1/3}$ respectively. The crossover point (${\lambda }_c$) is 
indicated by the vertical arrow. The inset shows that the probability $P_0$ of 
agents with zero money also first disappears at almost the same point 
(${\lambda }_c$) indicated by the vertical arrow.\\ 
\noindent
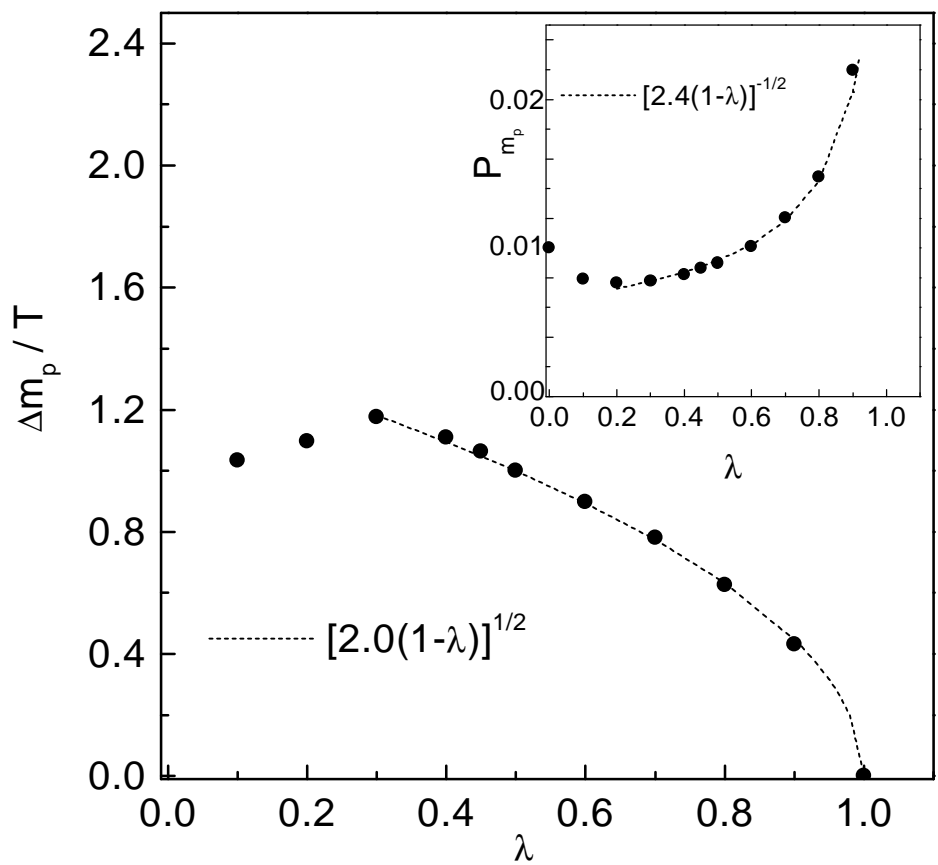
{\bf Fig. 4} : The variation of the distribution half-width $\Delta m_p$ as 
function of $\lambda$. The dashed curve corresponds to 
$\sqrt {2.0(1-\lambda )}$. The inset shows the peak height $P_{m_p}$ as 
function of $\lambda$. The dotted curve here corresponds to 
$1/\sqrt {2.4(1-\lambda )}$.\\ 
\noindent
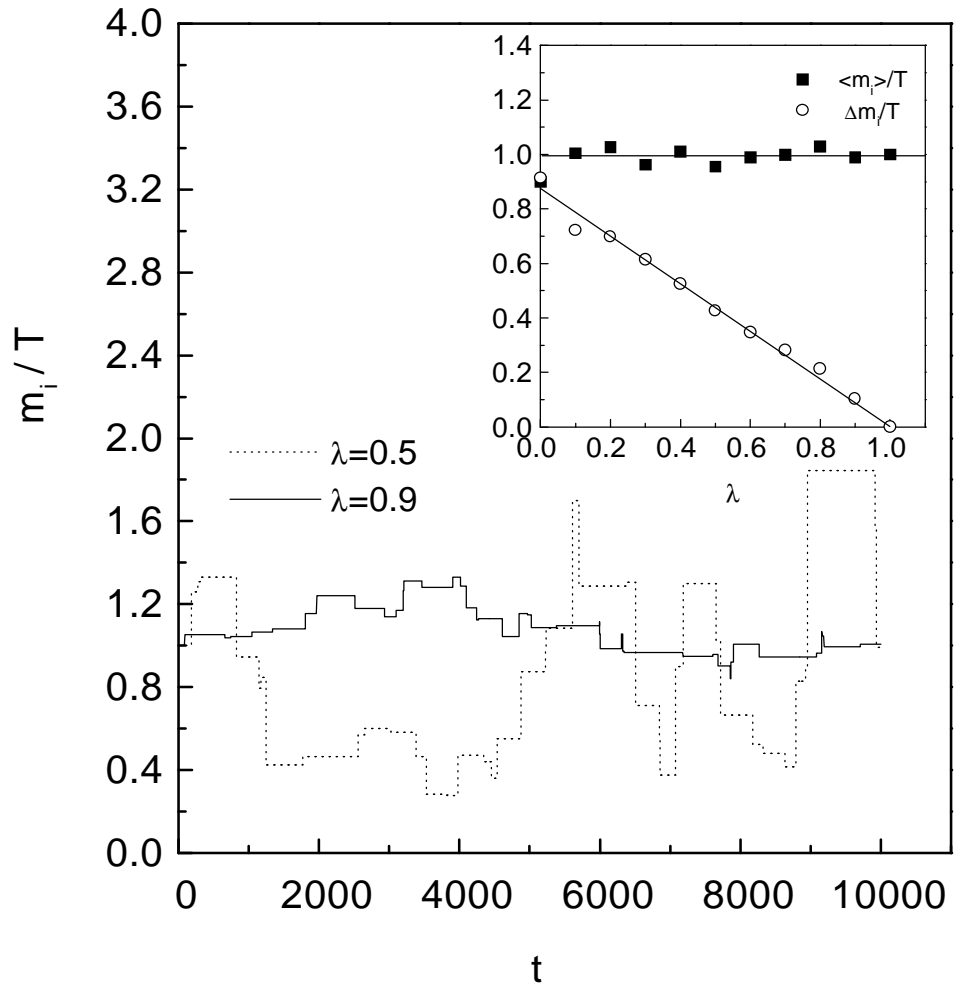
{\bf Fig. 5} : Time variation of money of an individual agent $m_i$ for two 
different values of  $\lambda$. The inset shows the variations of  
$\langle m_i \rangle$ and its fluctuations $\Delta m_i$ as functions of 
$\lambda$.\\ 
\end{document}